\documentclass{elsart}
\usepackage[all]{xy}
\usepackage{graphicx}
\journal{Physics Letters B}
\begin{document}
%%%%%%%%%%%%%%%%%%%%%%%%%%%%%%%%%%%%%%%%%%%%%%%%%%%%%%%%%%%%%%
\def\be{\begin{equation}}
\def\ee{\end{equation}}
\def\bea{\begin{eqnarray}}
\def\beas{\begin{eqnarray*}}
\def\eea{\end{eqnarray}}
\def\eeas{\end{eqnarray*}}
\def\ba{\begin{array}}
\def\ea{\end{array}}

%%%%%%%%%%%%%%%%%%%%%%%%%%%%%%%%%%%%%%%%%%%%%%%%%%%%%%%%%%%%%%%
%
\begin{frontmatter}
\title{Dynamically Induced Spontaneous Symmetry Breaking in 3-3-1 Models}
\author[label1]{Alex G. Dias}
\author[label2]{C.A. de S. Pires }
\author[label3]{V. Pleitez}
\author[label2]{and P.S. Rodrigues da Silva}
\address[label1]{Instituto de  F\'{\i}sica, Universidade
de S\~ao Paulo,\\ Caixa Postal 66.318, 05315-970,S\~ao Paulo-SP, Brazil}
\address[label2]{Departamento de F\'{\i}sica, Universidade Federal da
Para\'{\i}ba, Caixa Postal 5008, 58059-970, Jo\~ao Pessoa - PB,
Brazil}
\address[label3]{Instituto de F\'{\i}sica Te\'orica, Universidade
Estadual Paulista \\
Rua Pamplona, 145, 01405-900 S\~ao Paulo, SP - Brazil}

\begin{abstract}
We show that in $SU(3)_C\otimes SU(3)_L\otimes U(1)_N$ (3-3-1) models embedded with a singlet scalar playing the role of the axion, after imposing scale invariance, dynamical symmetry breaking of Peccei-Quinn symmetry occurs through the one-loop effective potential for the singlet field. We, then, analyze the structure of spontaneous symmetry breaking by studying the new scalar potential for the model, and verify that electroweak symmetry breaking is tightly connected to the 3-3-1 breaking by the strong constraints among their vacuum expectation values.
This offers a valuable guide to write down the correct pattern of symmetry breaking for multi-scalar theories. We also obtained that the accompanying  massive pseudo-scalar, instead of acquiring mass of order of Peccei-Quinn scale as we would expect, develops a mass at a much lower scale, a consequence solely of the dynamical breaking. 
\end{abstract}


\end{frontmatter}

\date{\today}

\maketitle

%%%%%%%%%%%%%%%%%%%%%%%%%%%%%%%%%%%%%%%%%%%%%%%%%%%%%%%%%%%%%%%%%%%%%%%%%%%%%%%%%%%
\section{Introduction}
\label{sec:intro}

The origin of mass has been  one of the greatest mysteries in Particle Physics.  
The simplest way we envision particles getting their masses is through Higgs mechanism, which takes place when a gauge invariant theory undergoes spontaneous symmetry breaking (SSB) as its potential develops a non-trivial value (a non zero vacuum expectation value (VEV)) for the scalar field at its minimum.

In the context of the Standard Model (SM) this is achieved by means of the most general gauge invariant and renormalizable potential, $V=\mu^2\phi^\dagger\phi+\lambda(\phi^\dagger\phi)^2$. Stability forces us to choose $\lambda>0$. But the choice of the unique dimensional parameter in the model, $\mu^2$, is enforced to accommodate the mass generation mechanism. This ad hoc construction brings to the theory an 
undetermined parameter making impossible to directly predict the scalar particle mass.
However, thirty years ago, it was shown that quantum corrections can lead to SSB in a scale invariant theory through the computation of its effective potential, the Coleman-Weinberg (CW) mechanism~\cite{colwein}. In such theories a typical mass scale is connected to the dimensionless parameters through the so called dimensional transmutation, which trades a coupling constant by a mass parameter. So, the particle masses, resulting from the scalar field condensation, would have a dynamical origin.

At one loop level, the dominant contribution arising from the most massive particles leads to the following form for the SM effective potential, according to the CW mechanism~\cite{SH89},
\begin{equation}
V=B_{SM}\phi^4[\ln(\frac{\phi}{\langle \phi \rangle})-\frac{1}{2}].
\end{equation}
The coefficient $B_{SM}$  is given by 
\begin{equation}
B_{SM}=\frac{1}{64\pi^2\langle \phi \rangle^4}(3M_Z^4+6M_W^4-12M_t^4),
\label{BSM}
\end{equation}
\noindent
where $M_Z$, $M_W$ and $M_t$ are the neutral, charged gauge bosons and top quark masses, respectively.  
In order to have $V$ bounded from below, the condition $B_{SM}>0$ has to be verified.       
In this way, the Higgs mass could be predicted in terms of the known particle masses of SM. 
At the time this idea was proposed, it was only possible to put an upper bound on the Higgs mass, since the top quark was not detected yet and there was a lot of uncertainty on the possible value of its mass. However, with the top quark mass determination, $M_t\approx 174$ $GeV$, the CW mechanism seemed then  
definitely discarded in the context of the SM once it implied $B_{SM}<0$. Moreover, Eq.~(\ref{BSM}) was derived by considering that the Higgs self-coupling was negligible compared to the other couplings in the theory, an assumption which is valid when the Yukawa couplings are small. However, since we already know that top quark is heavy, the Eq.~(\ref{BSM}) loses its meaning as it stands, and should be properly modified in order to account for a large Higgs self-coupling~\footnote{Recently, Elias et. al, in the Ref.~\cite{elias}, have shown that renormalization group improved Coleman-Weinberg mechanism in SM predicts a Higgs mass around 216~GeV when leading logarithms are summed over.}. By the other side, some simple extensions of SM were studied by adding more scalar fields, allowing for a stable effective potential even for fermion masses above $M_t$~\cite{Inoue}. 
Also, one should care about finite temperature effects as addressed in Refs.~\cite{SH89,Flores,SherPRD}, but they are important only if fermions are involved in the computation of the effective potential, which is not going to be the case in the study we perform here.

In this work we consider the CW mechanism and its effects, in 3-3-1 models~\cite{331} endowed with a singlet scalar field playing the role of the axion~\cite{axion1,axion2}. These 3-3-1 models are very attractive extensions of the gauge sector of SM, possessing additional scalar multiplets and, among several nice features, they can naturally accommodate a Peccei-Quinn (PQ) symmetry~\cite{pal}, $U(1)_{PQ}$, and solve the strong-CP problem with an invisible axion stable under gravitational effects if an appropriate $Z_N$ symmetry is imposed~\cite{axion1,axion2}. Their scalar sector can have three~\cite{axion1} or five~\cite{axion2} neutral complex scalars, besides the singlet, which could develop VEV's. It is natural to ask if these  VEV's (or some of them) could be an outcome of some dynamically broken symmetry at a higher energy scale. We remark that throughout this text we are going to use the expression {\it dynamical symmetry breaking} as referring to CW mechanism leading to a nontrivial VEV, which should not be confused with some fermion condensation as, for instance, in Technicolor models.
Our aim then is to obtain SSB at the electroweak scale driven by dynamical breaking at this higher scale, specifically the PQ scale around $10^{12}$~GeV. 

It is suitable to stress that the method here developed can be useful in similar situations for models containing multiple scalars able to develop VEV's. Its application to 3-3-1 models contains all the information needed to closely follow the details of our procedure. We believe it is a valuable tool for model builders when dealing with several neutral scalars and facing the need to pick up non-trivial values among several VEV's.

Next we review the field content of the models imposing the scale invariance. Then we perform the one 
loop calculation of the effective potential and look for the possibility of obtaining SSB driven by dynamical breaking in model I and comment the results for model II, which are very similar. Finally, we present our conclusions.

\section{The models}
\label{sec:models}

The class of models we are going to deal here, known as 3-3-1, first proposed about ten years ago~\cite{331}, 
constitute a gauge extension of SM which, among several nice features, requires  a multiple of three fermion families for anomaly cancellation~\cite{331,pt,ppm}.  Also, QCD demands a maximum of sixteen fermions for asymptotic freedom, which in 3-3-1 models translate to the fact that the number of families cannot exceed five (if we believe that asymptotic freedom stays valid at 3-3-1 scale), automatically implying that only three families are allowed, naturally explaining the outstanding family problem.
Besides, there is a bunch of new particles and interactions which
make these models phenomenologically rich and attractive as an
alternative to the SM~\footnote{Electroweak models with $SU(3)$ symmetry also have some predictive
power concerning the observed value of the weak mixing angle and can be embedded
in theories of TeV-gravity (see Ref.~\cite{dimo} and references therein).}, since they can be tested already at next collider experiments aimed to work at the TeV scale.
Among the possible field representations, we choose to work with only three scalar triplets, which is the minimal scalar content needed to generate the mass spectrum in two of these models.
In the first model, an exotic lepton appears in the leptonic triplets and  right-handed neutrinos come in singlets. Although a restriction on its perturbative applicability was found in Ref.~\cite{alex1}, the situation can be improved by adding new fermions in non-fundamental representations~\cite{alex2}. 
The second model is the so-called 3-3-1 with right-handed neutrinos~\cite{ppm,v2}, where right handed neutrinos naturally belong to the lepton triplets, instead of singlets, avoiding the introduction of heavier leptons.

With a singlet scalar
$\phi\sim({\bf1},{\bf1},0)$ embedded in the model~\cite{axion1,axion2}, we are going to impose scale invariance so that no dimensional parameter is allowed at tree level. Without scale invariance, mass terms for the scalars are present in the lagrangian and the naturalness problem sets in, which means we have to fine tune, order by order, the smallness of these masses with the hugeness of Planck scale. Rather than explaining this fine tuning, we assume that the mass terms for the scalars are zero from the beginning, a choice which is no less natural than fine tuning, justifying scale invariance. 
Bellow we briefly review these models.
 
\subsection{Model I}
\label{sub1}
This version of 3-3-1 model contains exotic charged leptons and quarks, and can be minimally implemented by considering only three scalar triplets~\cite{pt}. Its lepton content transforms under the  gauge symmetry $SU(3)_C\otimes SU(3)_L\otimes U(1)_N$ as follows,
\begin{eqnarray}
&&f^{a}_L = \left (\nu^a_L \,\,\,\, l^a_L \,\,\,\, E^a_L \right )^T\sim({\bf 1},{\bf 3},0)\,, 
\nonumber \\ \nonumber \\
&& \nu_{aR}\sim({\bf 1},{\bf 1},0)\,,\,\,\,
l_{aR}\sim({\bf1},{\bf1},-1)\,,\,\,\,E_{aR}\sim
({\bf1},{\bf1},+1)\,.
\label{lepcarg}
\end{eqnarray}
with $a=1,2,3$ representing the three known generations. 
The left-handed quarks  transform as,
\bea 
Q_{iL} = \left(
d_{iL} \,\,\,\,
u_{iL} \,\,\,\,
j_{iL} \right )^T\,, && \,\,\,\,\,\,\,\,
Q_{3L} = \left(
u_{3L}\,\,\,\,
d_{3L}\,\,\,\,
J_{L}
\right )^T\,,
\label{quarks} \eea
where $Q_{iL}\sim({\bf 3},{\bf 3}^{*},- 1/3)$  and $Q_{3L} \sim ( {\bf 3},{\bf 3},2/3)$, while the right-handed ones transform as,
\begin{eqnarray}
 u_{a R}\sim({\bf 3},{\bf 1}, 2/3)\,,&&\,\,\,\,\, d_{a R}\sim({\bf 3},{\bf 1},-1/3)\,,\,\,\,
\nonumber \\ \nonumber \\
j_{mR}\sim({\bf 3},{\bf 1},-4/3)\,,&& \,\,\,\, J_{R}\sim({\bf 3},{\bf 1},5/3),
\label{singletquark}
\end{eqnarray}
with $i,m=1,\,2$ and $a = 1,\,2,\,3$.

In the scalar sector, this model possess three triplets responsible for the fermion masses,
\bea \chi &=& \left (
\chi^- \,\,\,\,
\chi^{--} \,\,\,\,
\chi^{0} \right )^T\,,\,\,\,
\eta = \left (
\eta^0 \,\,\,\,
\eta_1^- \,\,\,\,
\eta_2^{+} \right )^T\,,\,\,\, 
%\nonumber \\
\rho = \left (
\rho^+ \,\,\,\,
\rho^0 \,\,\,\,
\rho^{++}
\right )^T\,, \label{chieta}\eea
transforming as  $\chi \sim ({\bf 1},{\bf 3},-1)$, $\eta \sim ({\bf 1},{\bf 3},0)$ and $\rho \sim ({\bf 1},{\bf 3},1)$, respectively.
 
Since the number of independent fields is large enough, it was observed in Ref.~\cite{axion1} that a $Z_{13}$ symmetry can be naturally associated to this model. 
We can then write the most general, renormalizable, gauge, $Z_{13}$ and scale invariant potential for this model,
{\bea &&V = \lambda_1\eta^4
+\lambda_2\rho^4 +\lambda_3\chi^4 
%\nonumber \\ &&
+\lambda_4(\eta^{\dagger}\eta)(\rho^{\dagger}\rho)
+\lambda_5(\eta^{\dagger}\eta)(\chi^{\dagger}\chi)
+\lambda_6(\rho^{\dagger}\rho)(\chi^{\dagger}\chi) 
\nonumber \\ &&
+ \lambda_7(\rho^{\dagger}\eta)(\eta^{\dagger}\rho)
+\lambda_8(\chi^{\dagger}\eta)(\eta^{\dagger}\chi)+\lambda_9(\chi^{\dagger}\rho)(\rho^{\dagger}\chi)
 + [\lambda_{10}\epsilon^{ijk}\eta_i\rho_j\chi_k\phi \,+ H.c.]
\nonumber \\ && 
+(\phi\phi^*)\left[\lambda_{\phi\chi}(\chi^{\dagger}\chi)
+\lambda_{\phi\rho}(\rho^{\dagger}\rho) 
%\nonumber \\ && 
+ \lambda_{\phi\eta}(\eta^{\dagger}\eta)+
\lambda_\phi(\phi\phi^*)\right]
\,.
\label{V} \eea}

An additional symmetry can be easily recognized in the model, a $U_{PQ}(1)$ symmetry, which after spontaneous breaking implies an invisible axion~\cite{axion1}, since it is predominantly formed by the singlet field and couples directly only with exotic quarks.  This not only  solves the strong-CP problem through an invisible axion but the existence of a $Z_{13}$ symmetry guarantees the stability of this solution under gravitational effects. 

\subsection{Model II}
\label{sub2}

The 3-3-1 model with right handed neutrinos on a triplet representation was introduced in Ref.~\cite{ppm,v2}. 
This 3-3-1 model differs from the above one basically by its matter content, since it contains no exotic charges for fermions and the right handed neutrino already belongs to the triplet. Besides, its scalar sector also comes in three triplets but it has a different content, as described below, 
\bea f^a_L = \left (
\nu^a_L \,\,\,\,
e^a_L \,\,\,\,
(\nu^{c})_L^a
\right )^T\sim({\bf 1},{\bf 3},-1/3)\,,\,\,\,e_{aR}\,\sim({\bf 1},{\bf 1},-1), \eea
where, again, $a = 1,\,2,\, 3$ label the three families. In the quark sector, one generation comes in the triplet
fundamental representation of $SU(3)_L$ and the other two compose
an anti-triplet with the following content,
\bea 
Q_{iL} = \left(
d_{iL} \,\,\,\,
-u_{iL} \,\,\,\,
d^{\prime}_{iL} \right )^T\,, && \,\,\,\,\,\,\,\,
Q_{3L} = \left(
u_{3L}\,\,\,\,
d_{3L}\,\,\,\,
u^{\prime}_{3L}
\right )^T\,,
\label{quarks2} \eea
where $Q_{iL} \sim({\bf 3},{\bf 3^*},0)$  and $Q_{3L} \sim({\bf 3},{\bf 3},1/3)$, and the right-handed quarks transform as,

\bea  
u_{a R}\,\sim({\bf 3},{\bf 1},2/3)\,, && \,\,\,\, d_{a R}\,\sim({\bf 3},{\bf 1},-1/3)\,,\, 
\nonumber \\ \nonumber \\
\,\,d^{\prime}_{iR}\,\sim({\bf 3},{\bf 1},-1/3)\,, && \,\,\,\, u^{\prime}_{3R}\,\sim({\bf 3},{\bf 1},2/3),
\label{quarks3} \eea
with $i=1,2$. The primed quarks
are the exotic ones but with the usual electric charges.

In order to generate the masses for the gauge bosons and fermions,
the model requires only  three triplets of scalars, namely,
\bea 
\chi &=& \left(
\chi^0 \,\,\,\,
\chi^{-} \,\,\,\,
\chi^{\prime 0}
\right )\,, \,\,\,\, \eta = \left(
\eta^0 \,\,\,\,
\eta^- \,\,\,\,
\eta^{\prime 0}
\right )\,,\,\,\,\,
%\nonumber \\
\rho = \left(
\rho^+ \,\,\,\,
\rho^0 \,\,\,\,
\rho^{\prime +}
\right )\,. 
\label{chieta2}
\eea
with $\eta$ and $\chi$ both transforming as $({\bf 1},{\bf 3},-1/3)$
and $\rho$ transforming as $({\bf 1},{\bf 3},2/3)$.

As in model I, we can write
the most general scalar potential, invariant
under the gauge symmetry, a discrete $Z_{11}\otimes Z_2$~\cite{axion2} and also scale symmetry. Its form is exactly the same as the one in Eq.~(\ref{V}), except for the differences in the scalar triplets content. As before, a $U_{PQ}(1)$ symmetry is identified and as a result of its spontaneous breaking an invisible and stable axion emerges~\cite{axion2}.

In the next section we are going to present the effective potential for the singlet scalar field, $\phi$, studying the possibility of driving SSB at scales lower than PQ scale in model I. We also comment about the results in model II.

\section{Induced spontaneous symmetry breaking}
\label{sec:efectpot}
The usual procedure in developing the effective potential when dealing with multiple scalars is described in Ref.~\cite{SH89,SherPRD,gildener}. There the quantum corrections to the effective potential are computed considering the scalar fields altogether. This is done in such a way that guarantees the perturbative validity of the scheme throughout the computation, by choosing a specific direction in field space such that all coupling constants are kept small. What we are going to do here is something different, since the question we wish to answer is whether the dynamical breaking for just one field, the axion in our approach, can induce spontaneous breaking of 3-3-1 as well as the electroweak symmetries. This means that we have only to compute the quantum corrections for the singlet field and check if it leads to a stable effective potential that can trigger nonzero VEV's for some of the remaining neutral scalars of the model, namely, $v_\chi$, $v_\rho$  and $v_\eta$ which are in charge of producing the desired pattern of symmetry breakdown. 

Applying the CW mechanism by considering only the singlet condensation,
we computed the effective potential for the 3-3-1 models above presented. We then obtain the following renormalized effective potential:
\begin{eqnarray}
V_{eff}=B_{v_\phi^2}\phi_c^4\left[\ln\left(\frac{\phi_c^2}{v_\phi^2}\right)-\frac{1}{2}\right]\,,
\label{Vefr}
\end{eqnarray}
where $\phi_c^2 \equiv \phi^*\phi$, $v_\phi \equiv \langle \phi_c \rangle$ and the coefficient $B_{v_\phi^2}$ is given by,
\begin{eqnarray}
B_{v_\phi^2}=\frac{3}{128\pi^2}\left[(\lambda_{\phi\eta}^2 + \lambda_{\phi\rho}^2 + \lambda_{\phi\chi}^2)\right]\,.
\label{cofB}
\end{eqnarray}
Observe that there is no trace of $\lambda_\phi$ in this equation, since the renormalization procedure, along with the minimum condition, lead to a $\lambda_\phi\propto \lambda_{\phi i}^2$, where $i=\eta, \rho, \chi$, which are small couplings as we will see ahead, allowing to neglect $\lambda_\phi$ in Eq.~(\ref{cofB}).
With this result we can already be certain that this effective potential breaks the $U_{PQ}(1)$ symmetry at $v_\phi$ scale, which we associate to the PQ scale. It is clearly a stable effective potential since all terms appearing in $B_{v_\phi^2}$ are definite positive.

In order to address the problem of driving spontaneous breaking by the dynamical one, we have to analyze the constraint equations coming from the full potential. This means that we are going to substitute the $\lambda_\phi \phi^4$ in Eq.~(\ref{V}), by the term obtained in Eq.~(\ref{Vefr})
and check which, if any, non-trivial VEV's for the neutral scalars are consistent with a minimum. It should be remarked that in this sense we first considered the potential in Eq.~(\ref{V}) and obtained a nontrivial VEV for the singlet field, which we called dynamical breaking. Only then we are going to formulate constraint equations for the VEV's for the remaining scalars, knowing that $v_\phi$ is guaranteed to be nonzero by CW mechanism. However, now the potential is not given by Eq.~(\ref{V}), but by that potential changed after the dynamical breaking as pointed above.
As far as we know, this approach is new and could prove very useful in establishing the correct pattern of breaking for a multi-Higgs theory, which is our case for the models presented in Sec.~\ref{sec:models}.

At this point the two models should differ, at least in their neutral scalar content, leading to different constraint equations. 
We are going to develop the formalism for model I here, just mentioning the results for model II. According to 
Sec.~\ref{sub1}, in order to get its constraint equations we first make the shift in each neutral scalar field by its respective VEV, and plug them into the new potential, we obtain the following constraints:
\begin{eqnarray}
&& \left[2\lambda_1v_\eta^2 +\lambda_4v^2_\rho+\lambda_5v^2_\chi+\lambda_{\phi\eta} 
v_\phi^2\right]v_\eta+
\lambda_{10}v_\rho v_\chi v_\phi = 0\,,
\nonumber \\
&& \left[2\lambda_2v_\rho^2+\lambda_4 v_\eta^2+\lambda_6 v_\chi^2+\lambda_{\phi\rho} 
v_\phi^2\right]v_\rho+
\lambda_{10}v_\eta v_\chi v_\phi = 0\,,
\nonumber \\ 
&& \left[2\lambda_3v_\chi^2+\lambda_5  v_\eta^2+
\lambda_6  v_\rho^2+\lambda_{\phi\chi}  v_\phi^2\right]v_\chi+\lambda_{10} v_\eta v_\rho v_\phi = 0\,, \nonumber \\&& \left[\lambda_{\phi\eta} v_\eta^2 + 
\lambda_{\phi\rho} v_\rho^2+ \lambda_{\phi\chi} v_\chi^2\right]v_\phi+\lambda_{10} v_\eta v_\rho v_\chi = 0\,,
\label{ce}
\end{eqnarray}
where we have taken, $\langle \phi_c \rangle = \frac{1}{\sqrt2}v_{\phi}$, 
$\langle \chi^0 \rangle = \frac{1}{\sqrt2}v_{\chi}$, 
$\langle \eta^0 \rangle  = \frac{1}{\sqrt2}v_{\eta}$ and 
$\langle \rho^0 \rangle  = \frac{1}{\sqrt2}v_{\rho}$.

Notice that in these equations, the only VEV which is guaranteed to be nonzero is $v_\phi$, by the CW mechanism developed above for the $\phi$ field. When we ask if we can induce SSB at low energies by the dynamical one obtained for $\phi$ at PQ scale, we are actually asking if the constraint equations are consistent with nontrivial values for the remaining VEV's at TeV, the typical scale for breaking 3-3-1, and electroweak scales, otherwise we would have been in trouble since the known low energy physics could not be recovered with these models.
For model I, looking at the set of Eqs.~(\ref{ce}), we see that if one of the triplet VEV's is different from zero only nontrivial solutions are possible for the remaining ones. This is easy to see by initially imposing that some of them are null and checking the constraint consistency condition. Hence, if one admits that the couplings in the potential are non-zero (since we have no underlying symmetry implying null couplings), one gets to the conclusion that SSB is a non negotiable outcome. 

We also observe that the constraint equations, Eq.~(\ref{ce}), allow us to determine the coupling constants in terms of the remaining couplings and VEV's:
\begin{eqnarray}
\lambda_{\phi\eta} &=& \frac{-\lambda_1 v_\eta^4 + \lambda_2 v_\rho^4 + \lambda_3 v_\chi^4 + \lambda_6 v_\rho^2 v_\chi^2}
{v_\eta^2 v_\phi^2}
\nonumber \\
\lambda_{\phi\rho} &=& \frac{\lambda_1 v_\eta^4 - \lambda_2 v_\rho^4 + \lambda_3 v_\chi^4 + \lambda_5 v_\eta^2 v_\chi^2}
{v_\rho^2 v_\phi^2}
\nonumber \\
\lambda_{\phi\chi} &=& \frac{\lambda_1 v_\eta^4 + \lambda_2 v_\rho^4 - \lambda_3 v_\chi^4 + \lambda_4 v_\eta^2 v_\rho^2}
{v_\chi^2 v_\phi^2}
\nonumber \\
\lambda_{10} &=& -\frac{v_\eta^2 (\lambda_1 v_\eta^2 + \lambda_4 v_\rho^2) + v_\rho^2(\lambda_2 v_\rho^2 + \lambda_6 v_\chi^2) + v_\chi^2 (\lambda_3 v_\chi^2 + \lambda_5 v_\eta^2)}
{v_\eta v_\rho v_\chi v_\phi}\,,
\label{coup}
\end{eqnarray}
which shows that the singlet interaction with the triplets are suppressed by factors of $(v_\chi / v_\phi)$ and $(v_\chi^2 / v_\phi^2)$, since $v_\phi$ is much bigger than $v_\chi$, which is about a few TeV's.

We could go a step further and ask: ``what if the highest scale below $v_\phi$, which means $v_{\chi}$, is also dynamically non-zero?'', we would have to follow the above analysis again and verify the induced spontaneous breaking. 
The problem of considering radiative symmetry breaking with multiple scalars was already considered before (see second article in Ref.~\cite{Inoue}), but here we want this dynamical breaking to be only a part of the whole mechanism of SSB.
To proceed with this proposal we remark that, differently from the singlet field, the triplet component $\chi$ couples not only with scalars but with fermions and vector bosons too.
In this way we could expect to have the same problem of instability as in the SM, since now we have heavy exotic quarks and vector bileptons which acquire mass at $v_\chi$ scale. It would make no difference then if,
instead of making assumptions over the masses of these fields, we just assume that $v_\chi$ is nonzero from the beginning. Observe that this assumption is not worse than that we had in the model without CW mechanism, where besides assuming a nonzero VEV we also had additional free parameters, the mass scales in the scalar potential. In essence, we can argue that our method allows us to eliminate some of the free parameters of 3-3-1 through scale invariance, and this is a step forward in reducing the unknowns in the theory. The fact that we added a scalar singlet into the model has increased the number of parameters though, but it happened in exactly such an amount that we ended up with the same number of parameters  as before, where we had no singlet scalar and no scale invariance. This is an interesting result, since we are able to solve strong-CP problem without any additional cost of increasing the number of parameters.

As for model II, the analysis follows exactly in the same way, but things are not so straightforward as in model I. The additional complication comes from the fact that here we have more VEV's at hand, namely, $v_{\chi^\prime}$ and $v_{\eta^\prime}$. Some of the VEV's can remain zero and still keep consistency, $v_{\chi}=0$ and $v_{\eta^\prime}=0$ (in this case we would recover the constraints of model I, Eq.~(\ref{ce})), or $v_{\chi^\prime}=0$ and $v_{\eta}=0$, for instance. However, this only points to interesting possibilities with this model, since the second solution would imply spontaneous breaking of lepton number, leading to a triplet majoron~\cite{maj}, while the first solution would have no such a feature. We would have no way to select between these two possibilities by the method here exposed but, again, we would know which allowed pattern of breaking are consistent for this model.

Finally, we mention that when no dynamical breaking is considered, and the VEV's are assumed to be non-trivial, the massive pseudo-scalar, $A_0$, acquires a mass which is dominated by a $v_\phi$ factor, $M_{A_0}\approx\sqrt{\left|\lambda_{10}\right| v_\phi v_\chi}$. However, in our dynamical scheme, this does not happen and the dominant contribution to $A_0$ mass is given approximately by, 
\begin{eqnarray}
	M_{A_0} & \approx & \sqrt{\frac{\left|\lambda_{3}\right|}{2}(\frac{1}{v_\eta^2}+\frac{1}{v_\rho^2}) v_\chi^4}\sim {\mathcal O}({\mbox TeV})\,,
\end{eqnarray}
a result which can be rather interesting phenomenologically, since this scalar can be produced at energy scales around TeV instead of much higher energies as in the case before. Besides, it shows that it is not necessarily true that scalars embedded in such a dynamical context are forced to get mass at the breaking scale. This could have interesting implications for models involving scalars originated at such high energy scales.

The important lesson from this study is the fact that we have SSB driven by dynamical breaking and, in the models here presented, this happens in such a way that there is coherence between their structure and phenomenological aspects. For instance, it could be that some of the low energy VEV's, like $v_\eta$ or $v_\rho$, would not be simultaneously different from zero for the models studied, implying that the scheme for fermion mass generation in these models would need revision. 
It is clear that what differentiates the view above presented from applying CW mechanism directly to multi-Higgs models is a subtlety. We are using the mechanism to ignite SSB while caring for its consistency. Also, it is a whole new way of dealing with SSB, since this is not an ad hoc assumption of nontrivial VEV's, we are firmly basing its origin on the dynamically generated VEV.
We believe this method can serve as a guide for model builders when several scalar multiplets develop VEV's, since they should obey some inherent consistency conditions, as shown above. Moreover, it seems that this approach provides a way of choosing the right pattern of breaking. In our case, model II offers the possibility of having two of the VEV's identically zero, though we could also pick up one or both of them as nonzero if needed. 
Conversely, we could not arbitrarily impose zero values to VEV's as we wish, which is patently obvious in both models in what concerns $v_\rho$. As we have seen, since $v_\chi\neq 0$, there is no freedom to assign a null value to this VEV, and we would have been in trouble if for some reason we needed that.

\section{Conclusion}
\label{sec:con}

We have shown that two models of an electroweak model based in a $SU(3)_L\otimes U(1)_N$  symmetry, with three scalar triplets and a singlet axion, are able to accommodate dynamical symmetry breaking at high energies and trigger spontaneous breaking at lower scales, which was done considering scale invariance.

By analyzing the effective potential for the singlet scalar field we obtained  dynamical breaking of Peccei-Quinn symmetry through CW mechanism. This breaking, along with the assumption that the next highest VEV is nontrivial~\footnote{As discussed in the text, this VEV could also be generated dynamically depending on the mass relation among the fermions and bosons of the referred models.}, lead to effective potential constraint equations for the remaining VEV's that fixed the pattern of spontaneous  breaking. This is what we have called {\it dynamically induced spontaneous symmetry breaking}. It should be remarked that the spectrum of Goldstones in the scale invariant theory would not match the required one if no singlet were added, because without an
interaction among the three triplets there would be an additional global symmetry. However, this interaction is reintroduced disguised in quartic terms involving the singlet, stressing the importance of the singlet in this approach.

It is clear that we are not trying to explain the hierarchy of VEV's in such models, or proposing any new mechanism to obtain the masses dynamically. However, we are offering a new way of viewing SSB as something with dynamical origin, though keeping some aspects of usual spontaneous breaking. It would be useless if no advantage were incremented.
In general, the virtue of such an approach is the fact that it allows us to reduce the number of free parameters in the multi-scalar models, at least by avoiding dimensional parameters in the lagrangian through scale invariance. 
This justified the inclusion of a singlet scalar with additional advantages without increasing the number of unknowns in the model.
Moreover, it is a powerful method 
to identify which pattern of breaking we can assume without running into contradictions, for instance, like imposing that some neutral scalars do not develop a VEV when the consistency between CW mechanism and the potential constraint equations demand nontrivial VEV's.
We believe that this way of facing the problem can help in several similar situations where multiple scalars can assume VEV's but no guide is available to get it without ambiguity.
Besides, it is possible that the structure of other multi-Higgs models would imply that CW mechanism can drive the desired pattern of breaking at once, by considering only that the first breaking at the highest energy is dynamical. In this sense the technique employed here can prove itself powerful and more appealing. 
Also, we observed that massive scalars generated in the context of dynamical breaking can have mass much lower than the breaking scale.
It would be interesting to have further tests of this approach applying it to other multi-Higgs models.

\vspace{1.0cm} \noindent {\bf Acknowledgments}

\noindent 
This work was supported by Funda\c{c}\~ao de Amparo \`a Pesquisa
do Estado de S\~ao Paulo (FAPESP) (AGD), Conselho Nacional de 
Desenvolvimento Cient\'{\i}fico e Tecnol\'gico (CNPq) (CASP,VP,PSRS) and Funda\c{c}\~ao de Apoio \`a Pesquisa do Estado da Para\'{\i}ba (FAPESQ) (CASP).

\end{document}